%% file: paper.tex
\documentclass[usenatbib,useAMS]{mn2e}

\usepackage{times}
\usepackage{rotating}
\usepackage{graphicx}
\usepackage{epstopdf}
\usepackage{amsmath}
\usepackage{amssymb}
\usepackage{bbold}
\usepackage{multirow}
 
\usepackage{xcolor}

\usepackage{float}

\input{./HEADER}

\begin{document}

\title[Augmented Lagrangian Perturbation Theory]{\centering{Cosmological Structure Formation\\ with\\ Augmented Lagrangian Perturbation Theory}}

\author[F.~S.~Kitaura and Steffen He{\ss}]{Francisco-Shu Kitaura\thanks{E-mail: kitaura@aip.de, Karl-Schwarzschild-fellow} and Steffen He{\ss}\\
Leibniz-Institut f\"ur Astrophysik Potsdam (AIP), An der Sternwarte 16, D-14482 Potsdam, Germany}

\maketitle

\input{./abstract}

\input{./main}

{\small
\bibliographystyle{mn2e}
\bibliography{lit}
}

\end{document}

%% file: HEADER.tex
\newcommand{\mbi}[1]{\mbox{\boldmath$#1$}}

\newcommand{\lsim}{\mbox{${\,\hbox{\hbox{$ < $}\kern -0.8em \lower 1.0ex\hbox{$\sim$}}\,}$}}
\newcommand{\gsim}{\mbox{${\,\hbox{\hbox{$ > $}\kern -0.8em \lower 1.0ex\hbox{$\sim$}}\,}$}}

\def\beqn{\vspace{2mm}
\begin{eqnarray}} 
\def\eeqn{\vspace{2mm} 
\end{eqnarray}}

 \def\la{\langle}

\newcommand{\be}{\begin{equation}}
\newcommand{\ee}{\end{equation}}
\newcommand{\ba}{\begin{eqnarray}}
\newcommand{\ea}{\end{eqnarray}}
\newcommand{\brr}{\begin{array}}
 
\newcommand{\err}{\end{array}}
\newcommand{\bc}{\begin{center}}
\newcommand{\ec}{\end{center}}

%% file: abstract.tex
\begin{abstract}

We present a new fast and efficient approach to model structure formation with  Augmented Lagrangian Perturbation Theory (ALPT). Our method is based on splitting the displacement field into a long and a short-range component. The long-range component is computed by second order LPT (2LPT). This approximation contains a tidal nonlocal and nonlinear term. Unfortunately, 2LPT fails on small scales due to severe shell crossing and a crude quadratic behaviour in the low density regime. The spherical collapse (SC) approximation  has been recently reported to correct for both effects by adding an ideal collapse truncation. However, this approach fails to reproduce the structures on large scales where it is significantly less correlated with the $N$-body result than 2LPT or linear LPT (the Zeldovich approximation). We propose to combine both approximations using for the short-range displacement field the SC solution.  A Gaussian filter with a smoothing radius $r_{\rm S}$  is used to separate between both regimes. 
We use the result of 25 dark matter only $N$-body simulations to benchmark at $z=0$ the different approximations: 1st, 2nd, 3rd order LPT, SC and our novel combined ALPT model. This comparison demonstrates that our method improves previous approximations at all scales showing $\sim$25\% and $\sim$75\% higher correlation than 2LPT with the $N$-body solution at $k=1$ and 2 $h$ Mpc$^{-1}$, respectively. We conduct a parameter study to determine the optimal range of smoothing radii and find that the maximum correlation is achieved with $r_{\rm S}=4-5$ $h^{-1}$ Mpc.  This structure formation approach could be used for various purposes, such as setting-up initial conditions for $N$-body simulations, generating mock galaxy catalogues, cosmic web analysis or for reconstructions of the primordial density fluctuations.

\end{abstract}


%% file: main.tex
\begin{figure*}
\includegraphics[width=16.cm]{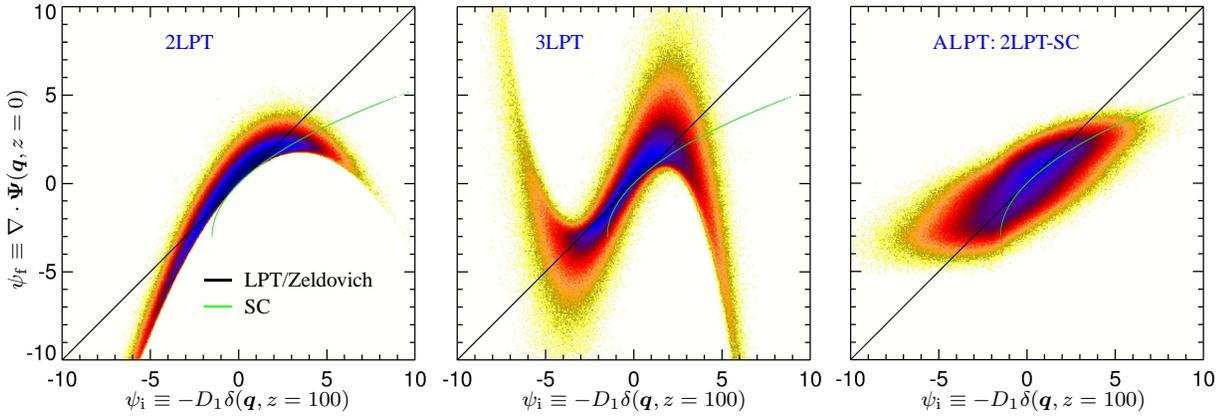}
\put(-400,130){\color{blue} 2LPT}
\put(-370,40){LPT/Zeldovich}
\put(-385,42.5){\color{black} \line(1,0){10} }
\put(-370,30){SC}
\put(-385,32.5){\color{green} \line(1,0){10} }
\put(-260,130){\color{blue} 3LPT}
\put(-110,130){\color{blue} \textsc{ALPT}: 2LPT-SC}
\put(-460,75){\rotatebox[]{90}{$\psi_{\rm f}\equiv\nabla\cdot\mbi\Psi(\mbi q,z=0)$}}
\put(-415,-5){$\psi_{\rm i}\equiv-D_1\delta(\mbi q,z=100)$}
\put(-265,-5){$\psi_{\rm i}\equiv-D_1\delta(\mbi q,z=100)$}
\put(-114,-5){$\psi_{\rm i}\equiv-D_1\delta(\mbi q,z=100)$}
\caption{
\label{fig:c2c} 
Cell-to-cell correlation between the linear initial overdensity field $D_1\delta(\mbi q,z=100)$ and the corresponding approximations for the divergence of the displacement field for the 10th realisation of our set  of simulations. The solid black line represents the LPT/Zeldovich approximation and the green curve the local SC model, which approximately fits the mean $N$-body relation. The nonlocal relations are given by the contours for various approximations: {\bf left panel: } 2LPT (quadratic relation). {\bf middle panel:} 3LPT (cubic relation) and {\bf right panel: } combined 2LPT-SC with $r_{\rm S}=4$ $h^{-1}$ Mpc. The dark colour-code indicates a high number and the light colour-code a low number of cells.}
\end{figure*}

\section{Introduction}

Numerical simulations are now an indispensable tool to study cosmological structure formation. The Poisson-Vlasov equations describing the gravitational interaction of matter in an expanding Universe,  are usually solved with a large number of test particles using so-called $N$-body codes \citep[for reviews on this subject see][]{1998ARA&A..36..599B,2012arXiv1209.5745K}. 
The most extended numerical techniques are the tree codes \citep[][]{1986Natur.324..446B} and the adaptive particle-mesh (PM) methods \citep[][]{1997ApJS..111...73K,2002A&A...385..337T}.
An example for a hybrid tree-PM code is \textsc{gadget} \citep[][]{gadget2} which uses the PM method to evaluate long-range forces and the tree method for short-range interactions. The forces between particles need to be recomputed in each time interval, requiring in total thousands of time steps to complete a single simulation.

Lagrangian Perturbation Theory (LPT) gives an approximate solution to structure formation by computing the displacement field from the primordial density fluctuations to the present Universe in one step \citep[for a review see][]{2002PhR...367....1B}. It has a wide range of applications in cosmology, giving an analytical understanding to structure formation \citep[][]{1970A&A.....5...84Z}.  It is commonly used to set-up initial conditions for numerical $N$-body simulations \citep[see][]{2006MNRAS.373..369C,jenkins10}. Another field of applications is the reconstruction of peculiar velocity fields \citep[see e.g.][]{2005ApJ...635L.113M,LMCTBS08,kitvel} and primordial density fluctuations  \citep[see e.g.][]{2006MNRAS.365..939M,2007ApJ...664..675E,kitlin,kigen}. It can also be used to change the cosmology of $N$-body simulations \citep[][]{2010MNRAS.405..143A}.
 The need for simulations to estimate uncertainties in cosmological parameter estimations from large scale structure surveys has raised the interest in approximate efficient structure formation models, which can be massively used. See \citet[][]{scocci,2002ApJ...564....8M,manera12} for generation of mock galaxy catalogues; \citet[][]{2012JCAP...04..013T} for modelling baryon acoustic oscillations; and \citet[][]{2011ApJ...737...11S} to increase the volume of a set of $N$-body simulations with smaller volumes. 
An improvement to linear LPT is given by second order LPT (2LPT) including nonlocal tidal field corrections. However, this is known to be a poor estimator in high and low density regions, being strongly limited by shell crossing \citep[][]{1996MNRAS.282..641S,2013MNRAS.428..141N}. 
Recently, local fits based on the spherical collapse model have been proposed \citep[][]{2006MNRAS.365..939M,2013MNRAS.428..141N}, which better match the mean stretching parameter (divergence of the displacement field) of $N$-body simulations. We propose in this work to combine the superiority of 2LPT on large scales with the more accurate treatment of the spherical collapse (SC) model on small scales including a collapse truncation of the stretching parameter, which acts as a viscosity term. Our algorithm splits the displacement field into a long-range and a short-range component, the first one being given by 2LPT and the second one by the truncated SC model. Both are combined by using a Gaussian filter with smoothing scales of 4-5 $h^{-1}$ Mpc radii, being this scale the only free parameter in our model.

This letter is structured as follows: in the next section (\S \ref{sec:theory}) we recap LPT and the SC model. In section \S \ref{sec:method} we present our method. We then show  (\S \ref{sec:results}) our numerical experiments comprising 250 simulations with various approximations. Finally (\S \ref{sec:conc}) we present our conclusions.

\begin{figure*}
\vspace{-1.75cm}
\includegraphics[width=8.cm]{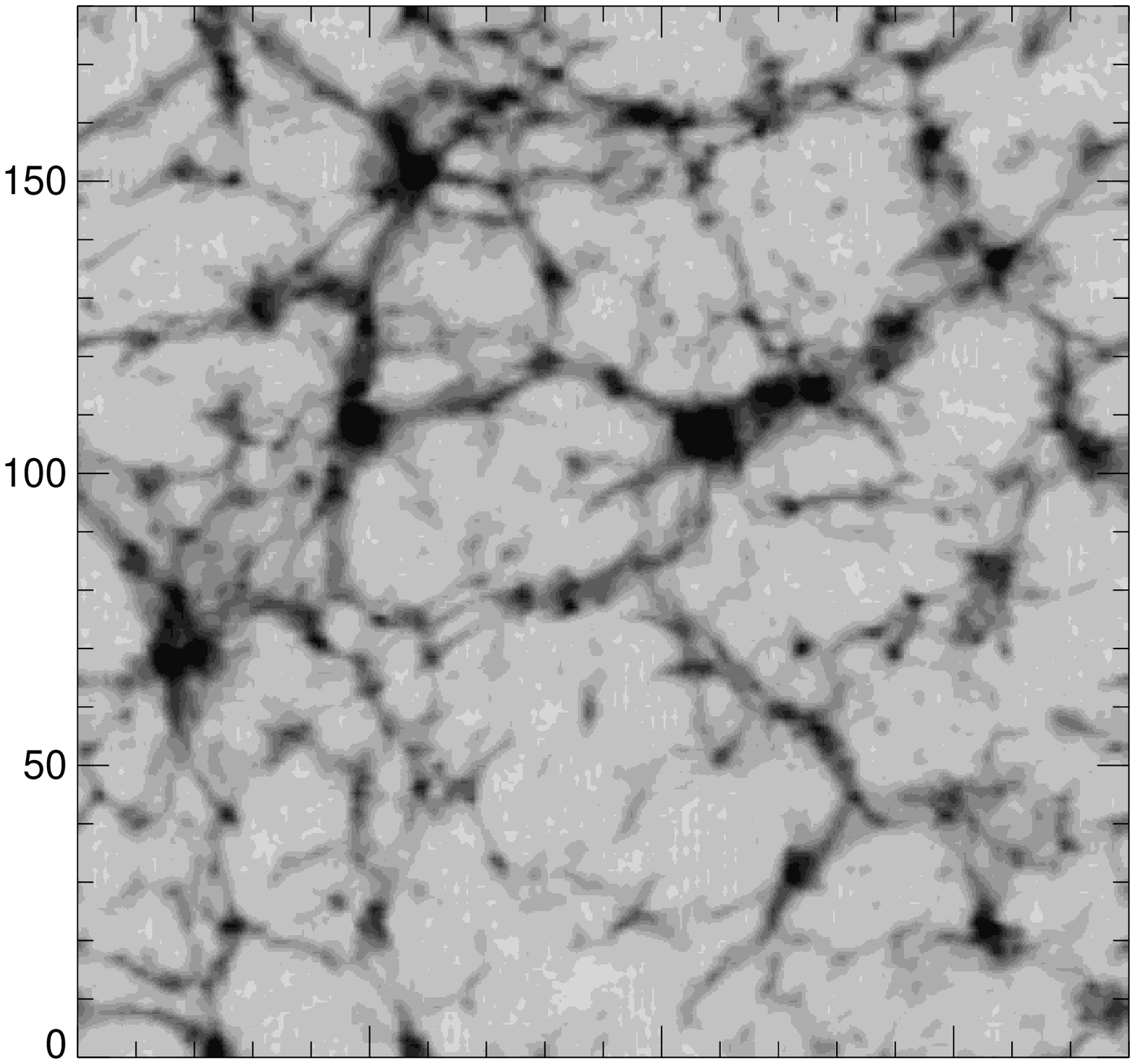}
\put(-230,150){\rotatebox[]{90}{{$Y$ [$h^{-1}$Mpc]}}}
\hspace{-1.05cm}
\includegraphics[width=8.cm]{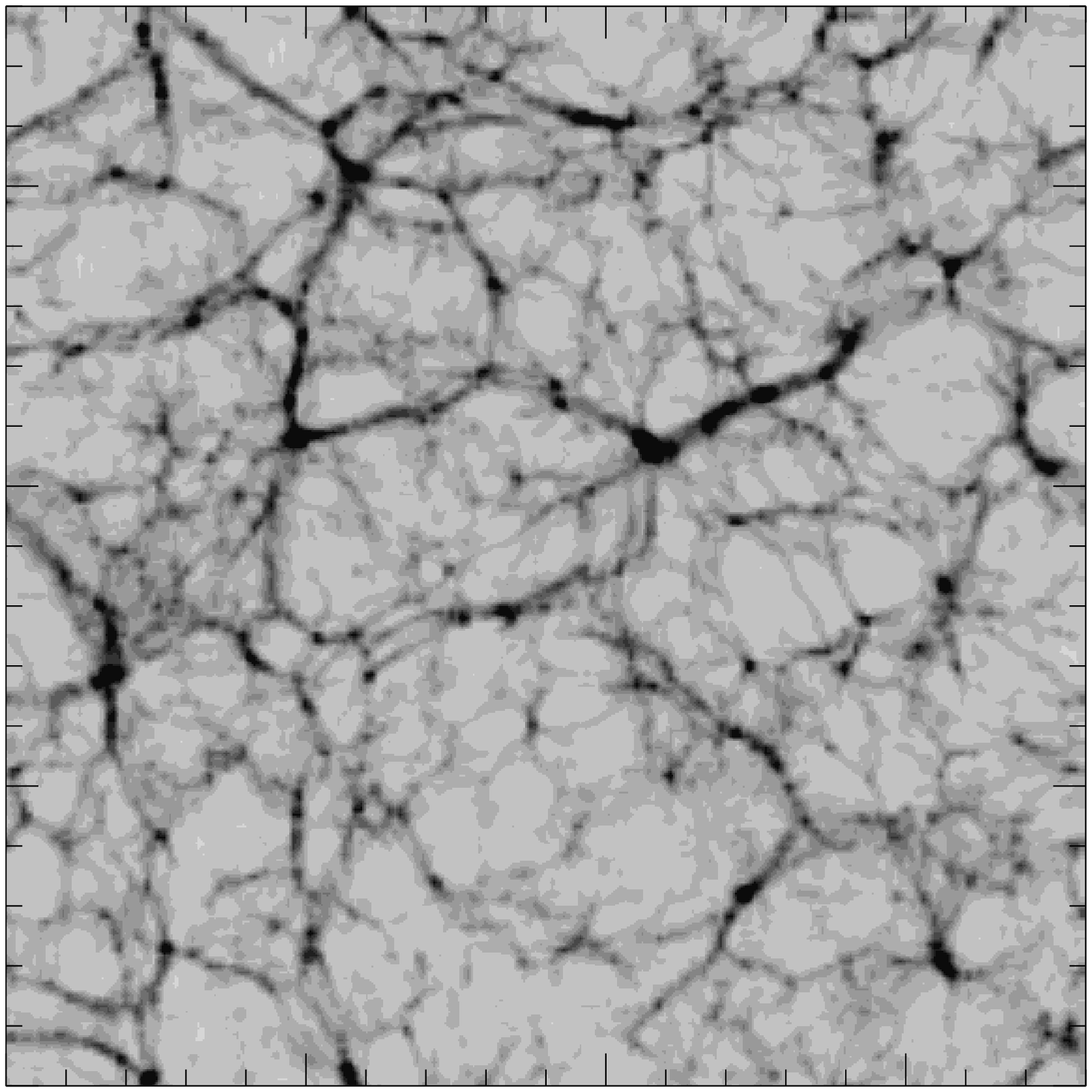}\\
\vspace{-3.27cm}
\includegraphics[width=8.cm]{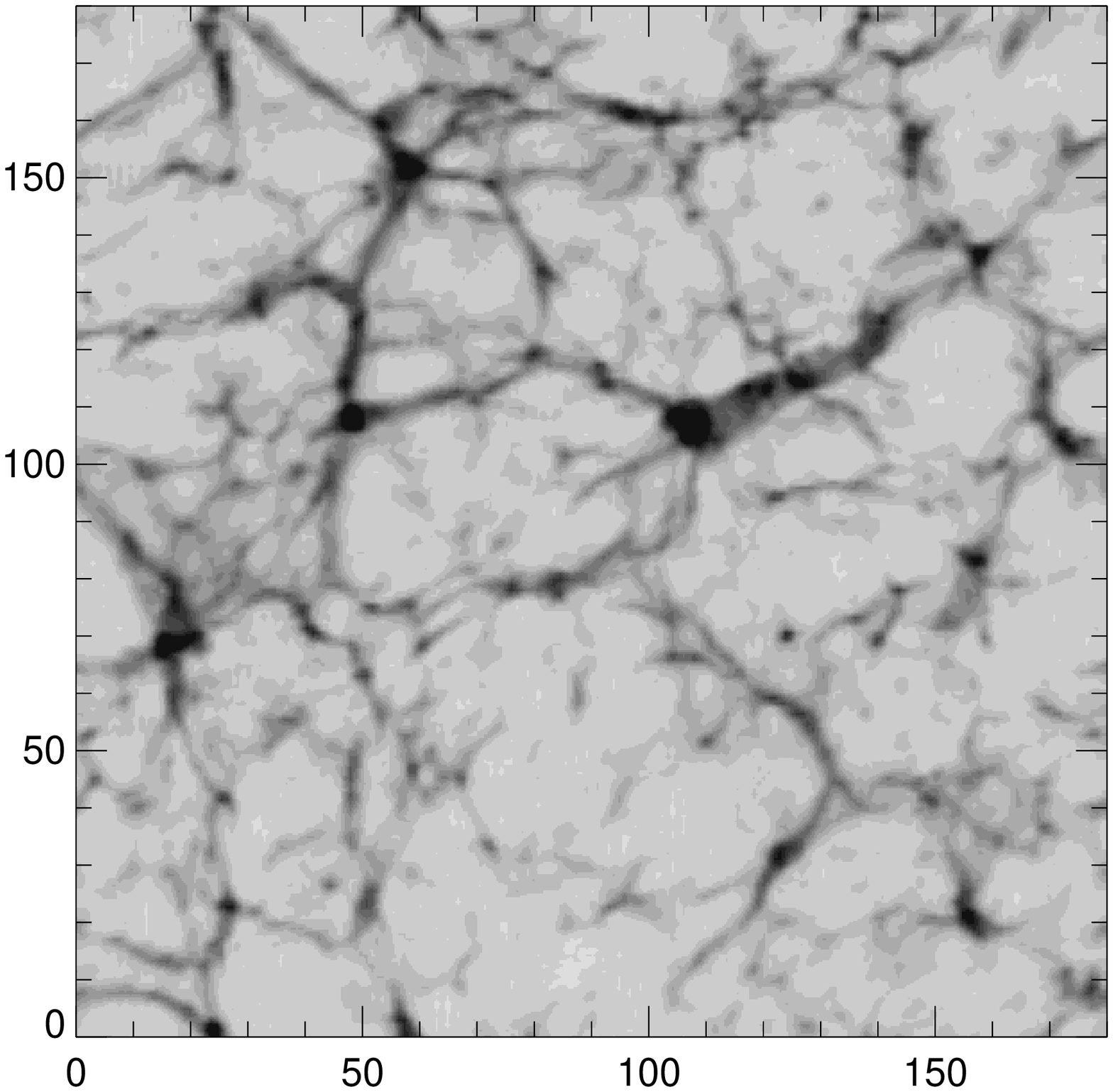}
\put(-120,27){{$X$ [$h^{-1}$Mpc]}}
\put(-230,150){\rotatebox[]{90}{{$Y$ [$h^{-1}$Mpc]}}}
\hspace{-1.05cm}
\includegraphics[width=8.cm]{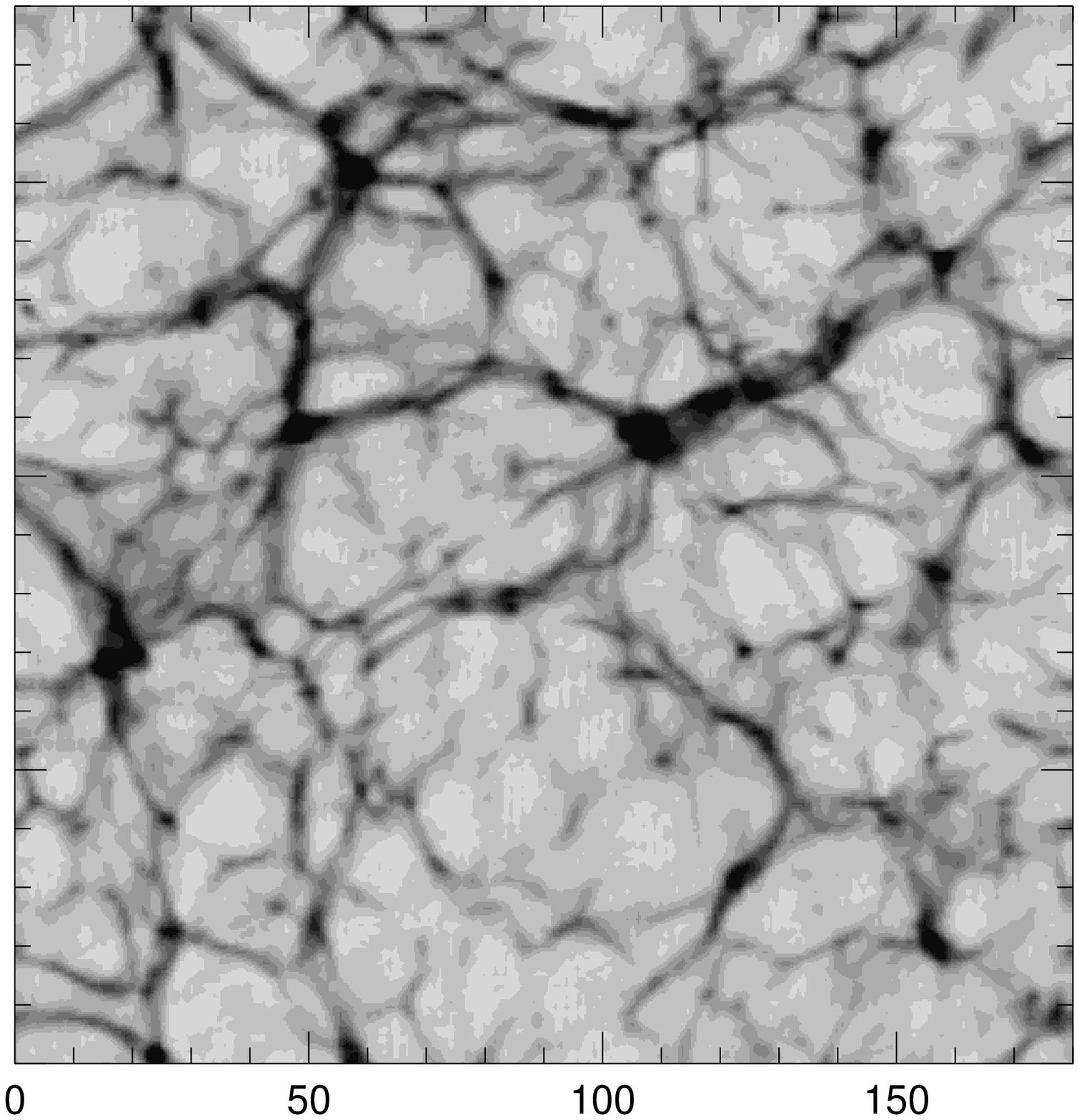}
\put(-120,27){{$X$ [$h^{-1}$Mpc]}}
\vspace{-1cm}
\caption{
\label{fig:sims} 
Slices for various structure formation models at $z=0$ of  $\sim$13 $h^{-1}$ Mpc thickness (averaged over 9 neighbouring slices) through a volume of 180 $h^{-1}$ Mpc per side on a mesh with $128^3$ cells (clouds-in-cells gridding). The logarithm of the density field ($\ln(2+\delta)$) is shown using the same colour range ($[-1,1.5]$) in all panels.  {\bf Upper left panel:}  2LPT; {\bf upper right panel:} $N$-body;  {\bf lower left panel:} combined 2LPT with 2LPT including collapse threshold (2LPT-2LPTC); {\bf lower right panel:}  combined 2LPT with SC model (2LPT-SC) for the 10th realisation of our set of simulations.
}
\end{figure*}

\section{Theory}

\label{sec:theory}

Let us define the positions of a set of test particles at an initial time $t_i$  by $\mbi q$ and call them the Lagrangian positions. The  final comoving positions $\mbi x$ (called Eulerian positions) corresponding to the same set of test particles at a final time $t_f$ are related to the Lagrangian positions $\mbi q$ through the displacement field, ${\mbi\Psi}$:
\begin{equation} 
\label{eq:lag}
 {\mbi x} = {\mbi q} + {\mbi\Psi} \, .
\end{equation}
Hence, the displacement field encodes the whole action of gravity during cosmic evolution.
An approximation is to consider that the displacement field is a function of the initial conditions only, and can be described by straight paths.
The various models consider the relation between the divergence of the displacement field and the linear initial field: $\psi=\psi(\delta^{(1)})\equiv\nabla\cdot\mbi \Psi(\delta^{(1)})$, where $\psi$ is the so-called stretching parameter. Let us call the previous equation the stretching parameter relation.
Lagrangian Perturbation Theory to third order yields the following expression for curl-free fields \citep[see][]{1994MNRAS.267..811B,bouchet1995,catelan}:
\ba
\label{eq:3lpt} 
 \psi_{\rm 3LPT}&\equiv&\nabla\cdot\mbi \Psi_{\rm 3LPT}\\
\hspace{0.cm}   &=&-D_1\delta^{(1)} + D_2\delta^{(2)}+D_{3{\rm a }}\delta^{(3)}_{{\rm a }}+D_{3{\rm b}}\delta^{(3)}_{{\rm b}}\,,\nonumber
\ea
 where $D_1$ is the linear growth factor, $D_2$ the second
order growth factor,  \{$D_{3{\rm a }},D_{3{\rm b}},D_{3{\rm c}}$\} are the 3rd order growth factors corresponding to the gradient of two scalar potentials ($\phi^{(3)}_{{\rm a }}$,$\phi^{(3)}_{{\rm b }}$). Particular expressions can be found in \citet[][]{bouchet1995} and \citet[][]{catelan}: $D_2=-3/7\, \Omega^{-1/143} D_1^2$, $D_{3{\rm a}}=-1/3\,\Omega^{-4/275}D_1^3$, $D_{3{\rm b}}=1/4 \cdot 10/21\,\Omega^{-269/17875}D_1^3$. 
The term $\delta^{(2)}({\mbi q})$ represents the
`second-order overdensity' and is related to the
linear overdensity field by the following quadratic expression:
\begin{equation}\label{twolpt_source}
 \delta^{(2)}({\mbi q})=\sum_{i>j} 
    \Big( \phi^{(1)}_{,ii}({\mbi q})\phi^{(1)}_{,jj}({\mbi q})-
    [\phi^{(1)}_{,ij}({\mbi q})]^2\Big),
\end{equation}
 The potentials $\phi^{(1)}$ and $\phi^{(2)}$ are obtained by
  solving a pair of Poisson equations:
  $\nabla^2_q\phi^{(1)}({\mbi q}) =  \delta^{(1)}({\mbi q})$,
where $\delta^{(1)}({\mbi q})$ is the linear overdensity, and 
  $\nabla^2_q\phi^{(2)}({\mbi q}) =  \delta^{(2)}({\mbi q})$.
The first term is cubic in the linear potential
\be
\delta^{(3)}_{{\rm a}}\equiv\mu^{(3)}(\phi^{(1)})=\det\left(\phi^{(1)}_{,ij}\right)\,,
\ee
and the second term is the interaction term between the first- and the second-order potentials:
\be
\label{eq:3lptdelta2}
 \delta^{(3)}_{{\rm b}}\equiv\mu^{(2)}(\phi^{(1)},\phi^{(2)})=\frac{1}{2}\sum_{i\ne j} 
    \Big( \phi^{(2)}_{,ii}\phi^{(1)}_{,jj}-
    \phi^{(2)}_{,ij}\phi^{(1)}_{,ji}\Big)\,.
\ee
Keeping terms only to first order is called the Zeldovich approximation \citep[][]{1970A&A.....5...84Z} and keeping terms to second order yields the 2LPT approximation. Fourth order LPT has been recently worked out for Einstein-de Sitter Universes \citep[see][]{Rampf:2012xa}.

Based on the nonlinear spherical collapse approximation, developed by \citet[][]{1994ApJ...427...51B}, \citet[][]{2006MNRAS.365..939M} found a local nonlinear expression for the stretching parameter relation:
 \ba
\label{eq:3lpt} 
 \psi_{\rm SC}\equiv\nabla\cdot\mbi \Psi_{\rm SC} & =&  3 \left[ \left(1-\frac{2}{3}D_1\delta^{(1)}\right)^{1/2}-1\right]\,.
\ea
This analytic formula has been recently found to fit very well the mean stretching parameter relation from an $N$-body simulation \citep[][]{2013MNRAS.428..141N}.  The stretching parameter cannot be smaller than $-3$ in this approximation, considering that for the ideal collapse to a point we have $\nabla\cdot\mbi x=0$ and thus $\psi=\nabla\cdot\mbi\Psi=-3$. We note the difference between using the relation for the comoving density field within the spherical collapse approximation \citep[][]{1994ApJ...427...51B}, which leads to infinite density in collapsed regions; and  the displacement field derived from that equation, which we use here to  suppress shell crossing and model low density regions \citep[][]{2006MNRAS.365..939M}.  We also note that  some theoretical studies have been done to improve LPT in the low density regime \citep[see][]{2011MNRAS.410.1454N}.

\section{Method}
\label{sec:method}

The method we present here is straightforward and computationally extremely fast.
Our ansatz is to expand the displacement field $\mbi\Psi(\mbi q,z)$ into a long-range $\mbi\Psi_{\rm L}(\mbi q,z)$ and a short-range component $\mbi\Psi_{\rm S}(\mbi q,z)$:
\be
\label{eq:dispsplit}
\mbi\Psi(\mbi q,z)=\mbi\Psi_{\rm L}(\mbi q,z)+\mbi\Psi_{\rm S}(\mbi q,z)\,.
\ee
We will rely on 2LPT for the long-range component: $\mbi\Psi_{\rm A}(\mbi q,z)=\mbi\Psi_{\rm 2LPT}(\mbi q,z)$. It is known that 2LPT fails towards small scales (see next section). Therefore, we propose to filter the displacement field resulting from the 2LPT approximation. We consider in this work a Gaussian filter ${\cal K}(\mbi q,r_{\rm S})$$=\exp{(-|\mbi q|^2/(2r_{\rm S}^2))}$, with $r_{\rm S}$ being the smoothing radius.
\be
\label{eq:dispsplit2}
\mbi\Psi_{\rm L}(\mbi q,z)={\cal K}(\mbi q,r_{\rm S}) \circ \mbi\Psi_{\rm A}(\mbi q,z)\,.
\ee
The spherical collapse approximation yields an extremely good local fit to the mean relation between initial and final stretching parameters. However, as it does not take into account the nonlocal tidal component it dramatically fails on large scales.  We will use this approximation to model the short-range component ($\mbi\Psi_{\rm B}(\mbi q,z)=\mbi\Psi_{\rm SC}(\mbi q,z)$):  
\be
\label{eq:dispsplit3}
\mbi\Psi_{\rm S}(\mbi q,z)=\mbi\Psi_{\rm B}(\mbi q,z)-{\cal K}(\mbi q,r_{\rm S}) \circ \mbi\Psi_{\rm B}(\mbi q,z)\,.
\ee
One may improve the method finding better approximations for the long-range $\mbi\Psi_{\rm A}$ or the short-range $\mbi\Psi_{\rm B}$ components.
We dub our method: Augmented Lagrangian Perturbation Theory \textsc{alpt}.

\section{Numerical experiments}
\label{sec:results}

\begin{figure}
\includegraphics[width=8.cm]{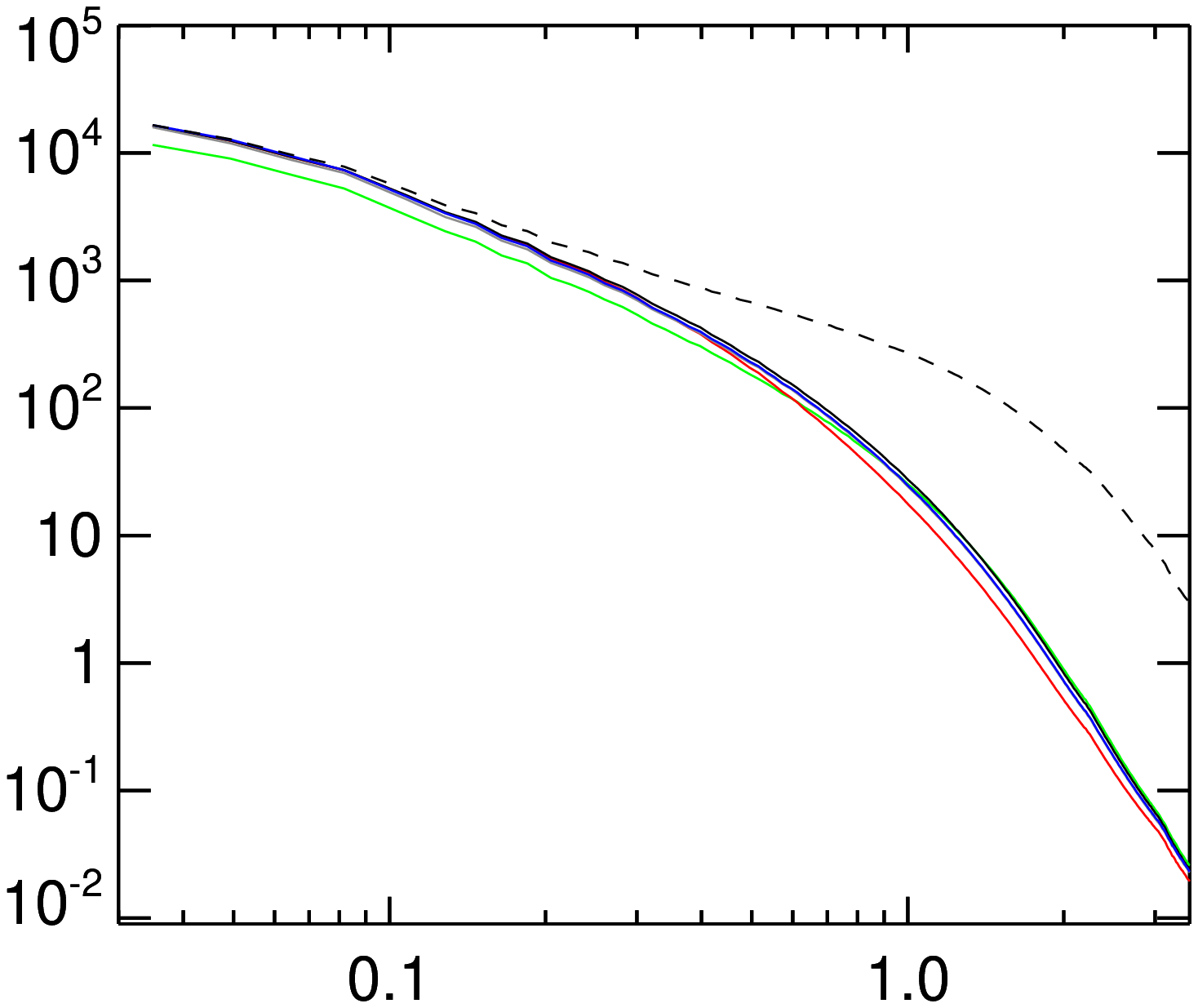}
\put(-235,110){\rotatebox[]{90}{$P(k)$ [$h^{-3}$ Mpc$^3$]}}
\put(-180,110){$N$ body}
\put(-195,112.5){\color{black} \line(1,0){4} }
\put(-189,112.5){\color{black} \line(1,0){4} }
\put(-180,100){2LPT-SC}
\put(-195,102.5){\color{black} \line(1,0){10} }
\put(-180,90){LPT/Zeldovich}
\put(-195,92.5){\color{blue} \line(1,0){10} }
\put(-180,80){2LPT}
\put(-195,82.5){\color{red} \line(1,0){10} }
\put(-180,70){3LPT}
\put(-195,72.5){\color{gray} \line(1,0){10} }
\put(-180,60){SC}
\put(-195,62.5){\color{green} \line(1,0){10} }
\put(-130,0){$k$ [$h$ Mpc$^{-1}$]}
\caption{
\label{fig:ps} 
Mean power-spectra for different structure formation models at $z=0$ using the same set of initial conditions, black dashed: $N$-body simulation; blue: LPT/Zeldovich; red: 2LPT;  gray: 3LPT; green: SC; black: combined 2LPT-SC approximation. The power-spectra include aliasing effects. 
}
\end{figure}

We benchmark different structure formation approximations by comparing them with the results from $N$-body simulations taking 25 randomly seeded initial conditions.
The simulations were done with \textsc{gadget-3}, an improved version of the publicly available cosmological code \textsc{gadget-2} \cite[last described in][]{gadget2}, using $384^3$ particles on a box with 180 $h^{-1}$ Mpc side. The initial conditions were generated using  the WMAP7 parameters \citep[][]{wmap7}. 
We consider different approximations of Lagrangian Perturbation Theory: 1st/Zeldovich, 2nd and curl-free 3rd  order LPT. We also use the spherical model (SC) \citep[][]{1994ApJ...427...51B,2006MNRAS.365..939M}.  Finally, we apply two improvements of the 2LPT model with $\mbi\Psi_{\rm A}(\mbi q,z)=\mbi\Psi_{\rm 2LPT}(\mbi q,z)$. In the first one we include the ideal collapse limit in the 2LPT approximation: $\mbi\Psi_{\rm B}(\mbi q,z)=\mbi\Psi_{\rm 2LPTC}(\mbi q,z)$. In the second one we additionally include the correction towards low densities using the spherical collapse model $\mbi\Psi_{\rm B}(\mbi q,z)=\mbi\Psi_{\rm SC}(\mbi q,z)$.
The relationship between the divergence of the displacement field and the linear overdensity field in various approximations is shown in Fig.~\ref{fig:c2c}. The linear behaviour of the Zeldovich approximation is represented by the black line with slope of 1. The quadratic and the cubic behaviours of 2nd and 3rd order LPT can be clearly seen in the scatter plots, left and middle panels, respectively. These were computed based on the 10th sample of our set of simulations. The SC model represented by the green curve serves as a good proxy for the mean relation corresponding to the $N$-body simulation \citep[see][]{2013MNRAS.428..141N}. The right panel shows the combined model 2LPT-SC which has been computed by taking the divergence of the resulting displacement field from Eq.~\ref{eq:dispsplit}.
This relation has several characteristics. The mean fits well the SC model towards low overdensities. We note that the SC underestimates the overdensities at low values of the stretching parameter. Our model reproduces the results of the $N$-body simulation better in this regime \citep[see Fig.~6 in][]{2013MNRAS.428..141N}. The stretching parameter truncates at stretching values of about 3-4 simulating the collapse and suppressing shell crossing. The 2LPT-2LPTC model is similar to the 2LPT-SC, however, retaining the quadratic behaviour of the 2LPT model towards underdense regions. 
We move $256^3$ particles forward in time using the different displacement fields presented here. Fig.~\ref{fig:sims} shows the results of the 2LPT models, classic and improved ones, as  compared to the $N$-body simulation for the 10th sample. One can clearly see the strong shell crossing of 2LPT and how this is suppressed in knots and thick filaments in both improved models. A careful inspection shows also that the low density regions are better captured by the 2LPT-SC model than by the 2LPT-2LPTC overdensity fields. It is remarkable how even small filaments which are present in the $N$-body simulation are also traced in the 2LPT-SC one.
Another important characteristic of the improved 2LPT models is that they preserve power on large scales as opposed to the SC one (Fig.~\ref{fig:ps} shows the mean power-spectra of the 25 realisations for each approximation). The improved 2LPT model has more power towards small scales than 2LPT showing that nonlinear structure formation is better captured.

\begin{figure*}
\includegraphics[width=8.cm]{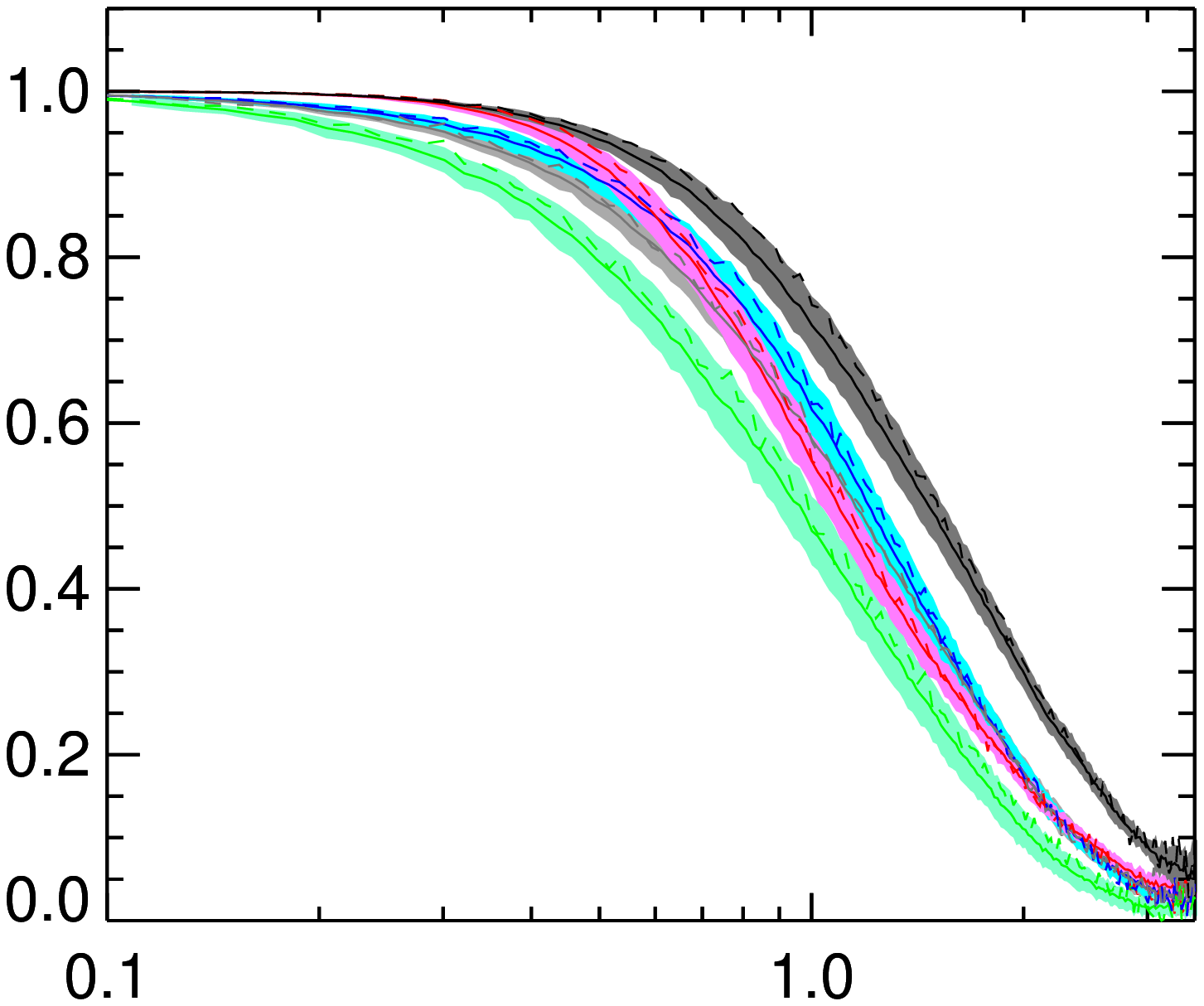}
\put(-180,30){$z=0$}
\put(-235,100){\rotatebox[]{90}{\rm $XP(k)$}}
\put(-180,100){ALPT: 2LPT-SC}
\put(-195,102.5){\color{black} \line(1,0){10} }
\put(-180,90){LPT/Zeldovich}
\put(-195,92.5){\color{blue} \line(1,0){10} }
\put(-180,80){2LPT}
\put(-195,82.5){\color{red} \line(1,0){10} }
\put(-180,70){3LPT}
\put(-195,72.5){\color{gray} \line(1,0){10} }
\put(-180,60){SC}
\put(-195,62.5){\color{green} \line(1,0){10} }
\put(-135,0){\rm $k$ [$h$ Mpc$^{-1}$]}
\hspace{-.5cm}
\includegraphics[width=8.cm]{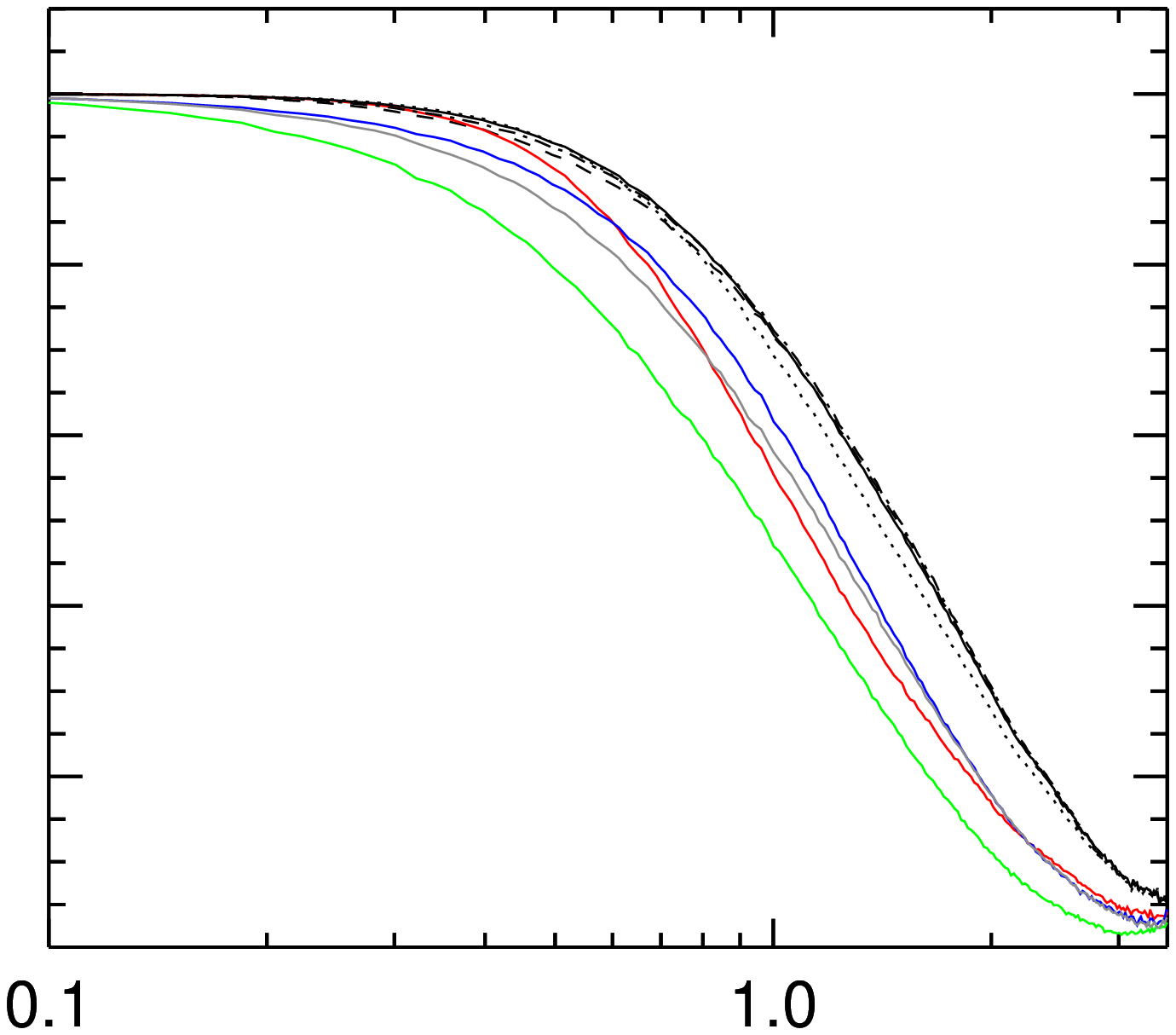}
\put(-180,30){$z=0$}
\put(-180,100){ALPT: 2LPT-SC}
\put(-180,90){$r_{\rm S}=3$ [$h^{-1}$ Mpc]}
\put(-195,90.5){$ \cdot \cdot \cdot$}
\put(-180,80){$r_{\rm S}=4$ [$h^{-1}$ Mpc]}
\put(-195,82.5){\color{black} \line(1,0){10} }
\put(-180,70){$r_{\rm S}=5$ [$h^{-1}$ Mpc]}
\put(-195,72.5){\color{black} \line(1,0){4} }
\put(-188,70.5){$\cdot$}
\put(-180,60){$r_{\rm S}=6$ [$h^{-1}$ Mpc]}
\put(-195,62.5){\color{black} \line(1,0){4} }
\put(-189,62.5){\color{black} \line(1,0){4} }
\put(-133,0){\rm $k$ [$h$ Mpc$^{-1}$]}
\caption{
\label{fig:xps} 
Normalised cross-power spectra $XP(k)\equiv\langle|\hat{\delta}_{\rm approx}(\mbi k)\overline{\hat{\delta}_{N{\rm body}}(\mbi k)}|\rangle/(\sqrt{P_{\rm approx}(k)}\sqrt{P_{N{\rm body}}(k)})$  between the matter overdensity from an $N$-body simulation $\delta_{N{\rm body}}$ and the corresponding approximation $\delta_{\rm approx}$: 1st/Zeldovich (blue), 2nd (red), 3rd (gray) LPT, SC (green) and the combined 2LPT-SC (black) model. {\bf On the left:} The mean cross-power spectra are given by the solid lines, the dashed line represents the 10th realisation of our set  of simulations, the 1 sigma contours are given by the corresponding light shaded regions. The 2LPT-SC model was computed with a threshold scale of $r_{\rm S}=4$ $h^{-1}$ Mpc. {\bf On the right:} same as left panel including the mean cross power-spectra for the following smoothing radii: $r_{\rm S}=3,4,5,6$ $h^{-1}$ Mpc (dotted, solid, dashed-dotted and dashed, respectively).
}
\end{figure*}

Finally, we assess the quality of our approach by computing the cross-power spectra between the various approximations and the $N$-body solution. We find that our approach yields a higher correlation on all scales by choosing the appropriate threshold scale (see Fig.~\ref{fig:xps}). We perform a parameter study to determine the optimal range of smoothing radii at $z=0$ and find that the maximum correlation is achieved with $r_{\rm S}=4-5$ $h^{-1}$ Mpc (see right panel in Fig.~\ref{fig:xps}). This result is in agreement with previous works which also found that 2LPT performs best down to scales of about $4-5$ $h^{-1}$ Mpc \citep[see][]{kitlin,kitvel,kigen}. When taking a smoothing scale of 6 $h^{-1}$ Mpc we start to get slightly worse results than 2LPT on very large scales $k<0.4$ $h$ Mpc$^{-1}$. We note that the 2LPT-2LPTC model yields similar cross-power spectra to the 2LPT-SC model, indicating that shell crossing in high density regions  reduces more dramatically the accuracy of 2LPT. We have checked that the 2LPT approximation with a collapse truncation does not improve the results of the SC model. In fact, including the collapse limit significantly degrades the cross-correlation with respect to the classical 2LPT model. At higher redshifts a smaller smoothing scale can be taken as the 2LPT solution becomes more accurate and the difference to the spherical collapse solution is reduced \citep[see Fig.~6 in][]{2013MNRAS.428..141N}. 

Furthermore, our results show that 3LPT does not improve the Zeldovich solution. We also find that the Zeldovich approximation performs better than 2LPT already at about $k=0.5-0.7$ $h$ Mpc$^{-1}$ depending on the realisation.

\section{Conclusions}

\label{sec:conc}

We have presented a new approach to model structure formation: Augmented Lagrangian Perturbation Theory (\textsc{alpt}). Our method is based on splitting the displacement field into a long- and a short-range component, given by 2LPT and the spherical collapse approximations, respectively. 
These are the advantages of our method:
\begin{itemize}
\item It fits very well the relation between the divergence of the displacement field and the linear overdensity field.
\item It considerably enhances the cross-correlation on all scales between the $N$-body solution with respect to previous approximations like the 1st (Zeldovich), 2nd, 3rd order Lagrangian Perturbation Theory and the spherical model approximation.
\item It requires only one additional parameter which defines the transition from the long-range to the short-range regime. We find an optimal choice of this parameter to be in the range 4-5 $h^{-1}$ Mpc for Gaussian smoothing.
\item Similar to 2LPT it preserves the power on large scales.
\item Since it is a one-step solver it is extremely fast and efficient as compared to $N$-body simulations.
\end{itemize}
The range of applications is very wide. Due to its improved modeling of structure formation this approach could be used for the following purposes:
\begin{itemize}
\item Set-up initial conditions for $N$-body simulations  \citep[see][]{2006MNRAS.373..369C,jenkins10}. 
\item Make mock galaxy catalogues \citep[see 2LPT based approaches][]{scocci,manera12,kitic}, which could be an improved tool for dark energy studies and cosmological parameter studies for present and future large-scale structure surveys.
\item Fast modeling of the Universe at high redshift to perform statistical analysis (e.g. Lyman-$\alpha$ forest, 21 cm line) \citep[see \textsc{21cmfast} which uses the Zeldovich approximation,][]{2011MNRAS.411..955M}.
\item To analyse the cosmic structure (cosmic web classification) and perform environmental studies \citep[][]{nuzaenv}.
\item For the inference of the primordial density fluctuations to perform baryon acoustic oscillation (BAO) reconstructions \citep[][]{2007ApJ...664..675E} or constrained simulations \citep[][]{klypin}. The method presented here can be directly implemented in \textsc{kigen} \citep[][]{kigen,kit2mrs} to recover the initial conditions to perform constrained simulations see \citet[][]{hesscs}.
\end{itemize}
We will address each of these issues in forthcoming publications.

{\bf Acknowledgements}
FSK is indebted to the Universidad Aut{\' o}noma de Madrid for the hospitality and thanks Gustavo Yepes and Francisco Prada for encouraging discussions. SH acknowledges support by the Deutsche Forschungsgemeinschaft under the grant $\rm{GO}563/21-1$. The work was carried out under the HPC-Europa2 project (project number: 228398) with the support of the European Commission - Capacities Area - Research Infrastructures.